# COMMUNICATION

## Nanoarchitectonics in fully printed perovskite solar cells with carbon-based electrodes

Dmitry Bogachuk,*[a] Jessica Girard,[b] Siddharth Tilala,[a] David Martineau,[b] Stephanie Narbey,[b] Anand Verma,[b] Andreas Hinsch,[a] Markus Kohlstädt [a,c] and Lukas Wagner [d]

**A sacrificial film of polystyrene nanoparticles was utilized to introduce nano-cavities in mesoporous metal oxide layers. This enabled the growth of larger perovskite crystals inside the oxide scaffold with significantly suppressed non-radiative recombination and improving device performance. This work exemplifies potential applications of such nanoarchitectonic approaches in perovskite opto-electronic devices.**

Perovskites solar cells (PSC) enjoyed the attention of numerous researchers over the recent years due to the fascinating nature of these materials. [1–3] After an unprecedentedly rapid increase in the reported power conversion efficiencies (PCE), nowadays, more attention is given within the perovskite community to the device stability.[4] The PSC with carbon-based back-electrodes (CPSC) are well-known to be highly stable under different ambient and operational conditions.[5,6] However, their PCEs are still lower compared to state-of-the-art PSCs with metal-based electrodes.[7]

Mesoscopic CPSCs consist of fluorine-doped tin oxide (FTO) front electrode, compact titanium dioxide (c-TiO₂), mesoporous titanium dioxide (m-TiO₂), zirconium dioxide (ZrO₂) and carbon back-electrode. The carbon electrode is usually treated at high temperatures, meaning that perovskite can only be introduced after the back-electrode is deposited. Consequently, the entire device structure must be porous, to allow the solution to fill the scaffold and to consequently crystallize perovskite within this porous matrix between the front and back electrode. Therefore, the perovskite crystal size and density of grain-boundaries (GBs) are determined by the pore size, because they confine perovskite growth. Since the particles of m-TiO₂, ZrO₂ layers and corresponding pores are typically 20 nm in diameter or less,[8,9] perovskite grown within these layers is expected to have numerous GBs with surface defect states. As has been shown in our earlier study,[10] this causes substantial non-radiative recombination and limits their potential to reach highest open-circuit voltage ($V_{OC}$).

Consequently, to overcome this limitation, one needs to enlarge the perovskite crystal size. This implies that the pores in which the perovskite is formed need to be enlarged in order to grow crystals with less GBs. In a CPSC stack there are three porous layers: m-TiO₂, ZrO₂ and carbon – all of which are filled with perovskite. The m-TiO₂ should be thin to minimize parasitic absorption and interfacial non-radiative recombination losses[8] as well as porous enough to avoid capacitance issues,[11] which arise from the high dielectric constant of TiO₂.[12] As a result, PSCs with TiO₂ having small surface area (e.g. having only a planar TiO₂) typically have severe hysteresis of the $JV$-characteristics.[1,4] This limits the possibility to tune the well-optimized pore size of the m-TiO₂ layer. The porosity of carbon layer affects the solution permeability[13] and the electrode series resistance[14] but not the non-radiative recombination since perovskite embedded into this layer is essentially "in the dark" (*i.e.* far beyond the optical generation profile[8]). In contrast, enlarging the pores of ZrO₂ is still a rather unexplored option and will likely suppress the non-radiative recombination rate of perovskite within due to reduced density of GBs and have the most prominent effect on the device performance.[10]

[a] *Fraunhofer Institute for Solar Energy Systems ISE, Heidenhofstr. 2, 79110 Freiburg, Germany.*
[b] *Solaronix SA, Rue de l'Ouriette 129, Aubonne 1170, Switzerland.*
[c] *Freiburg Materials Research Center FMF, University of Freiburg, Stefan-Meier-Str. 21, 79104 Freiburg, Germany.*
[d] *Solar Energy Conversion Group, Department of Physics, Philipps-University Marburg, Renthof 7, 35037 Marburg, Germany.*
† Footnotes relating to the title and/or authors should appear here.
Electronic Supplementary Information (ESI) available: See DOI: 10.1039/x0xx00000x

In this work we propose the use of nanoarchitectonics – utilization of nano-scaled and well-arranged structural units – to modify the architecture of an opto-electronic device, in this case a CPSC. By utilizing inkjet-printed polystyrene (PS) nanoparticles as a sacrificial template between m-TiO₂ and ZrO₂ layers we were able to facilitate the creation of cavities at this interface, which will be filled with large perovskite crystals after the cell manufacturing is complete. The developed inkjet-printing process allows to deposit highly ordered layers of hexagonally-arranged PS nanoparticles, on which ZrO₂ layer could be flawlessly deposited. During ZrO₂ sintering, the PS-nanoparticles pyrolyze, acting as a sacrificial template to nanostructure the ZrO₂ layer. A 3D-nanotomogramm of the sample obtained via scanning electron-microscope couple with focused ion-beam ablation (SEM-FIB) confirmed the successful alternation of the device structure and introduction of the enlarged cavity. Furthermore, steady-state and transient PL-measurements clearly show that perovskite-filled scaffolds with such cavity exhibit a reduced non-radiative recombination, slightly boosting the charge carrier lifetime and device $V_{OC}$. The nanoarchitectured CPSCs also exhibited a slightly better ideality factor and less non-radiative fill factor (FF) losses coming from higher pseudo-fill factor. Furthermore, the nano-architectured cells exhibit a small improvement in short-circuit current density ($J_{SC}$) by approximately 1.5 mA/cm² suggesting an increase in light absorption due to slightly higher photoabsorber volume. Overall, the champion nanoarchitectured cell had an impressive PCE of 15.7%. Although this study is focused particularly on improving the performance of CPSCs with such technique, the concept of nanoarchitectonics could be explored in other PSC architectures as well and provides an immense potential for further device enhancement.

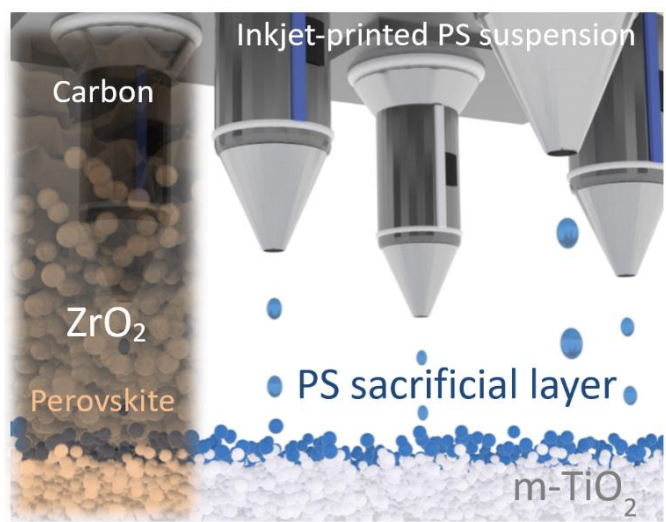

Figure 1: Conceptual illustration depicting an inkjet-printing process of sacrificial layer of polystyrene (PS) particles for nanoarchitecturing perovskite solar cell with carbon-based electrodes with solar cell structure depicted on the left.

Firstly, to be able to deposit a film of PS nanoparticles, a PS suspension was diluted with methanol and spin-coated on the FTO. After drying the film, its morphology was investigated via scanning electron microscope (SEM). We note that to improve the layer packing density, surfactant Triton X-100 was added, allowing for the deposition of a compact and homogeneous film (Fig. S1a). However, to be used in an up-scalable inkjet-printing process (Fig. 1), the diluted suspension needed to be optimized further. In order to avoid clogging of the head nozzles, methanol was replaced with isopropanol, and the resulting suspension was further diluted with 1-pentanol. Furthermore, we used formamide to induce better Marangoni flows, which counteract the convective flows, thus minimizing droplet diameter and avoiding the "coffee-ring effect".[15] With this optimized PS suspension, a well-ordered arrangement of PS-nanoparticles was obtained. Surprisingly, the PS particles in the inkjet-printed layers are hexagonally-arranged and appear to be even more ordered than in the case of spin-coated films (Fig. S1b). However, the spin-coated films were slightly thicker than the printed ones (Fig. S1c-d) which could be related to the difference in the ink composition. For further experiments, we only utilize the inkjet-printing technique to coat the substrate with PS nanoparticles.

ZrO$_2$ was then screen-printed on top of the deposited PS-layer followed by a slow sintering process (heating up to 500ºC), similarly to other nanoarchitectonic approaches to structure the photoabsorber in PSCs.[16] Such method resulted in the creation of localized nano-cavities between the FTO and the ZrO$_2$, without delamination of the latter layer from the substrate (Fig. 2a). For more visual comparison, Fig. 2b shows that the reference sample with ZrO$_2$ deposited on FTO substrate does not possess such feature. However, the cavity height is significantly lower than the thickness of the initially deposited nano PS layer. During ZrO$_2$ sintering PS particles undergo the transition from solid state to rubbery and viscous states, followed by their depolymerization and evaporation. Therefore, most likely the height of PS layer shrinks during the transition to the rubbery state.

To quantify the nano-cavity height distribution in a sample we utilize FIB-SEM tomography, whereby we successively polish the cross-section and record the SEM images of the polished surface allowing to obtain information about the sample's appearance along its depth. Combining these 2D SEM images along the axis perpendicular to the plane of taken images allows us to reconstruct a 3D image of our nanoengineered structure and compare it with the reference one (Fig. 2c). Then the 3D-nanotomogramms were analyzed (via image analysis software ImageJ) to plot a distribution of the nanocavity height in the polished ~2 µm³ sample volume. The distribution in Fig. 2c revealed that additional localized cavities with up to 100 nm in height have been introduced overall across the sample. In comparison, 3D-nanotomogramm and the corresponding distribution of a reference sample demonstrates that it does not possess such feature and most cavities are below 10 nm, likely to originate from the porosity of ZrO$_2$.

Next, we filled the nanoarchitectured samples with perovskite to study the effect of large perovskite grains grown within ZrO$_2$ on the radiative and non-radiative recombination rates. The mixed-dimensional 2D/3D perovskite photoabsorber layer was primarily composed of methylammonium lead iodide (MAPbI$_3$) with a small addition of 5-aminovaleric acid (5-AVA) similarly to our previous works.[6,10,14] As can be seen from the TRPL measurement shown in Fig. 3a, the effective decay time constant ($\tau$) of PS-samples was found to be 760 ns by fitting the PL behavior to a bi-exponential decay. This is significantly higher than in the case of perovskite embedded in the reference ZrO$_2$ layer, which was found to be 570 ns. Higher $\tau$ in nanoarchitectured samples suggests a prolonged charge-carrier lifetime, by effective suppression of non-radiative recombination via reduction of grain boundaries in perovskite.

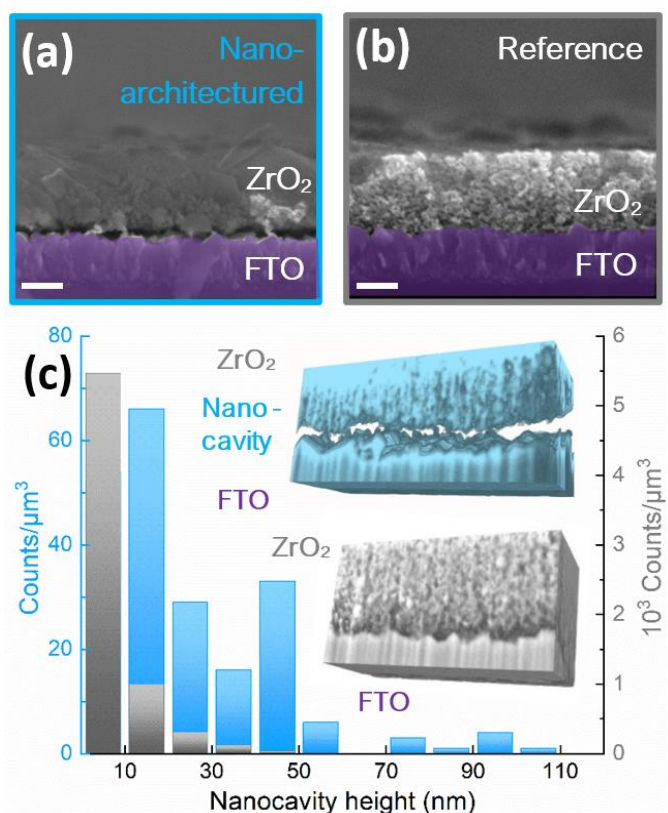

Figure 2: Cross-sectional scanning electron microscope (SEM) images of the (a) nanoarchitectured layer of ZrO$_2$ and (b) reference layer of ZrO$_2$. Scale bars – 500 nm. (c) Distribution of the cavity height in the nanoarchitectured (blue) and reference (grey) samples, obtained from 3D-nanotomography using focused ion beam (FIB) ablation.

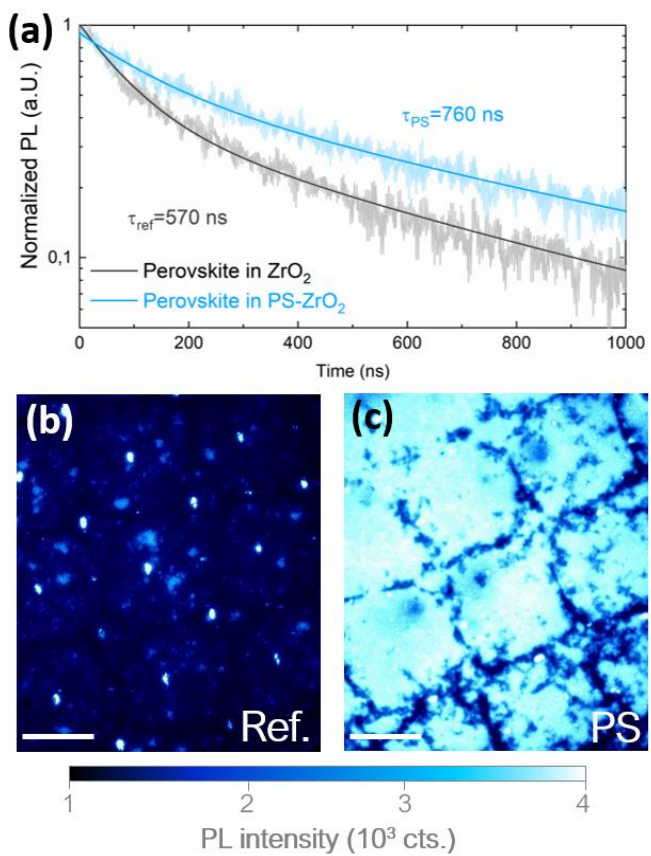

Figure 3: (a) Time-resolved photoluminescence (TRPL) measurements of perovskite embedded in the reference ZrO₂ and nanoarchitectured ZrO₂ (PS-ZrO₂), showing difference in the decay constant and charge carrier lifetime. Steady-state PL maps of (b) reference perovskite solar cells with carbon electrodes and (c) nanoarchitectured cells with PS-sacrificial layer. Scale bar – 150 µm.

To test whether this nanoarchitecturing approach also results in the improvement of PV devices, we manufactured CPSCs with and without the nano-cavity. In accordance with the prolongation of PL decay and charge carrier lifetime, the PL mapping in Fig. 3b reveals that the reference CPSCs have a much lower PL intensity than the PS samples (Fig. 3c). Note that the square patterns in both PL maps originates from the mesh of the screen-printing stencil (during deposition of TiO₂) and is not related to the PS layer. Therefore, high PL intensity and long charge carrier lifetime indicates that the non-radiative recombination in such cells is reduced, which would positively affect the device performance.

The measured statistical data obtained from current-voltage measurements of the manufactured cells is presented in Fig. 4a, demonstrating that the nanostructured samples cause a significant enhancement in $V_{OC}$ by approximately 30-40 mV and in $J_{SC}$ by ~1.5 mA/cm². According to the Ross relation[17] (detailed derivation can be found in Supplementary note 1) $V_{OC}$ can be expressed as a function of photoluminescence quantum yield (PLQY):

$$V_{OC} = V_{OC}^{rad} + \frac{k_B T}{q} \ln(PLQY) = V_{OC}^{rad} + \frac{k_B T}{q} \ln\left(\frac{j_{rad}}{j_{tot}}\right), \quad (1)$$

where $V_{OC}^{rad}$ denotes the potential difference between quasi-Fermi levels at the radiative limit, $k_B$ – Boltzmann constant, $T$ - temperature and $q$ is the elementary charge. The sum of radiative ($j_{rad}$) and non-radiative ($j_{nrad}$) recombination current densities yields the total recombination current density $j_{tot}$. Therefore, a difference in $V_{OC}$ directly depends on the difference in PLQY and $j_{nrad}$ via:

$$\Delta V_{OC} = V_{OC,1} - V_{OC,2} = \frac{k_B T}{q} \ln\left(\frac{j_{rad,1}}{j_{rad,2}}\right) = \frac{k_B T}{q} \ln\left(\frac{j_{tot} - j_{nrad,1}}{j_{tot} - j_{nrad,2}}\right) \quad (2)$$

Considering that the mean difference in $V_{OC}$ between reference and nanoarchitectured cells is ~33 mV, the $j_{rad}$ and PLQY should increase by 3.5 times, which agrees remarkably well with the intensity difference of the PL maps shown in Figs. 3b and 3c. We also have simulated CPSCs with different grain-sizes using 1D drift-diffusion model similarly to our previous work[10] to demonstrate an increase in QFLS and implied $V_{OC}$ due to grain size increase (Fig. S2). Notably the grain size starts to play particularly important role in cells with large grain boundary density due to logarithmic dependency of the QFLS and the measured increase in $V_{OC}$ by 33 mV would correspond to doubling of grain size from 25 to 50 nm (30.5 mV according to the simulated results).

From the $JV$-statistics in Fig. 4a we also highlight an exceptional reproducibility of this approach with a standard deviation of PCE of only 0.2%, in comparison to the deviation of 0.5% in the case of reference cells. Furthermore, the champion PS-engineered cell had a $V_{OC}$ of 943 mV and a PCE of 15.7% (measured from reverse scan).

To further investigate the reduction of non-radiative recombination losses we measured the light intensity-dependent $V_{OC}$ to extract the ideality factor $n$, which is often used to describe its non-ideal diode behaviour as:

$$V_{OC} = n\frac{k_B T}{q} \ln\left(\frac{J_{ph}}{J_0} + 1\right), \quad (3)$$

where $J_{ph}$ is photocurrent and $J_0$ is the dark saturation current density. By fitting the $V_{OC}$ measurements to equation (3), the slope determined by $n$ (since $\frac{k_B T}{q}$ is a constant) was found and shown in Fig. S3. Slightly lower ideality factor confirms that non-radiative recombination losses are suppressed in CPSCs with nanocavity.

Next, we used the light intensity-dependent behaviour of $V_{OC}$-$J_{SC}$ pairs to construct a so-called pseudo-$JV$ curve (p$JV$ curve). Since $V_{OC}$ is measured under open circuit, the charge transport outside of the cell is absent and therefore the effect of series resistance can be neglected. Thus, the p$JV$ curve represents a JV curve of a cell in the absence of charge transport losses, which is then only limited by non-radiative recombination losses.[18–20] Fig. 4b shows the $JV$- and p$JV$-curves of the reference and nanoarchitectured CPSCs based on the measurements in Fig. S3. It indicates that by minimizing the series resistance losses, the nanoarchitectured CPSC could reach a potential PCE (pPCE) of nearly 18 %. In addition, Fig. 4b shows a $JV$-curve of an "ideal" solar cell with this perovskite bandgap (1.61 eV),[14] where only radiative recombination is present, according to the Shockley-Queisser limit. By comparing the measured $JV$-curves with an ideal one, we see that the biggest PCE loss is due to the $V_{OC}$, which is reduced by nearly 400 mV. This substantial gap implies that although the presented approach can suppress the non-radiative recombination at the GBs to some extent, there is still a lot of room for further improvement.

In summary, we demonstrated an effective strategy to engineer the architecture of perovskite solar cells with carbon-based electrodes in order to create nanocavities between the m-TiO₂ and the ZrO₂ layers. These nanocavities were formed by depositing ZrO₂ on an inkjet-printed and highly-ordered layer of polystyrene nanoparticles, which acts as a sacrificial template to foster the controlled formation of a nanocavity. The success of this approach was confirmed by SEM measurements coupled with cross-sectional FIB polishing to produce a 3D-nanotomogram of the nanoarchitectured sample, from which we quantified the cavity height distribution by image analysis. Despite the fact that the cavity height (and therefore maximum perovskite crystal height) is <100 nm we demonstrate by transient and steady-state PL-mapping measurements that these nanocavities still cause a suppression of non-radiative recombination in perovskite photoabsorber and lead to an improvement in charge carrier lifetime. Consequently, the $V_{OC}$ has been improved by nearly 40 mV and accompanied by a small increase in $J_{SC}$, resulting in a champion PCE of 15.7 %. Furthermore, by performing light intensity dependent measurements, we show that the ideality factor $n$ has been reduced in cells with larger perovskite crystals, agreeing with the results from PL analysis. Based on these measurements, we construct a pseudo-$JV$ curve, demonstrating a potential of these nanoarchitectured cells to reach PCE of 17.9 %, if the charge-transport losses would be minimized. The findings of this work revolve around a conceptually new idea of nanoarchitectonics in PSCs as a new method to boost their PV performance (particularly in the case of devices with mesoporous structures), and potentially bring it closer to the radiative limit.

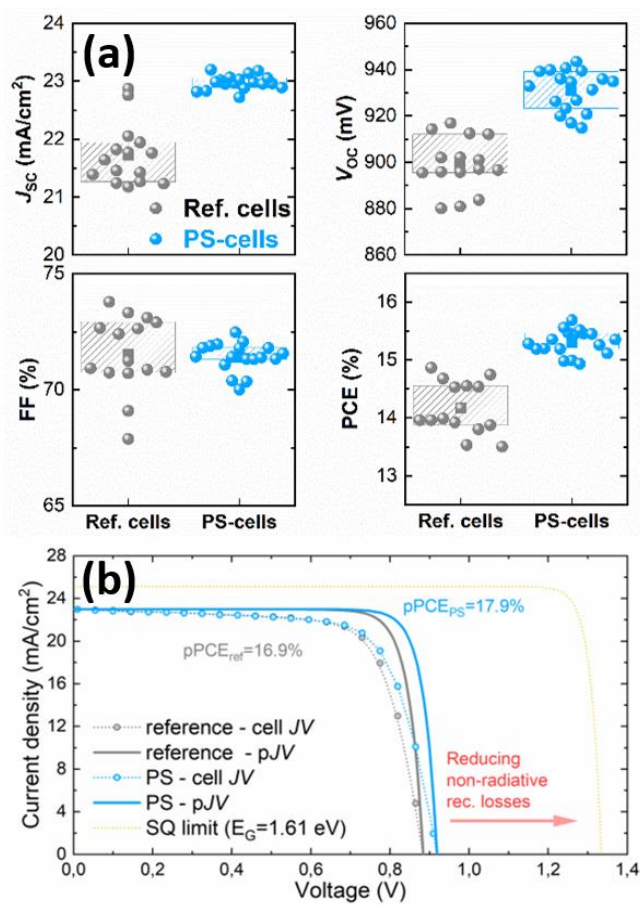

Figure 4: (a) Short-circuit current density ($J_{SC}$), open-circuit voltage ($V_{OC}$), fill factor (FF) and power conversion efficiencies of manufactured CPSCs with and without sacrificial PS layer. (b) $JV$-curves and pseudo-$JV$ curves of the champion reference and nanoarchitectured cells, in comparison with the curve at the Shockley-Queisser limit.

This work was funded by project UNIQUE supported under umbrella of SOLAR-ERA.NET_cofund by ANR, PtJ, MIUR, MINECO-AEI and SWEA, within the EU's HORIZON 2020 Research and Innovation Program (cofund ERA-NET Action No. 691664). And with the support of the German Federal Environmental Foundation (DBU).

## Conflicts of interest

There are no conflicts to declare

## Notes and references